\documentclass[11pt]{article}
\usepackage{latexsym}

\newcommand{\beq}{\begin{equation}}
\newcommand{\eeq}{\end{equation}}
\newcommand{\beqs}{\begin{eqnarray}}
\newcommand{\eeqs}{\end{eqnarray}}
\newcommand{\tr}{\mathrm{tr\,}}
\newcommand{\WW}{{\cal W}}
\newcommand{\refe}[1]{(\ref{#1})}
\newcommand{\nn} {\nonumber}
\newcommand{\weff}{W_{\mathrm{eff}}}
\newcommand{\wtree}{W_{\mathrm{tree}}}

\begin{document}
\begin{titlepage}
\vskip 2.5cm
\begin{center}
{\LARGE \bf Effective superpotential for $U(N)$ \\
with antisymmetric matter}
\vspace{2.71cm}

{\Large
Riccardo Argurio}
\vskip 0.5cm
{\large \it Physique Th\'eorique et Math\'ematique\\
and\\ 
International Solvay Institutes\\
\smallskip
Universit\'e Libre de Bruxelles, C.P. 231, 1050 Bruxelles, Belgium}
\vskip 0.5cm
{\tt email: Riccardo.Argurio@ulb.ac.be}
\end{center}
\vspace{2cm}
\begin{abstract}
We consider an ${\cal N}=1$ $U(N)$ gauge theory with matter
in the antisymmetric representation and its conjugate, 
with a tree level superpotential containing at least quartic interactions
for these fields. We obtain the effective glueball superpotential in 
the classically unbroken case, and show that it has a non-trivial
$N$-dependence which does not factorize. We also recover additional
contributions starting at order $S^N$ from the dynamics of $Sp(0)$ factors.
This can also be understood
by a precise map of this theory to an $Sp(2N-2)$ gauge theory with
antisymmetric matter.
\end{abstract}

\end{titlepage}

\section{Introduction}
Supersymmetry makes it possible to understand at least some aspects
of the exact effective dynamics of confining gauge theories. In particular, 
for ${\cal N}=1$ supersymmetric gauge theories, it is possible to
determine exactly the low-energy effective F-terms and thus analyze
the quantum vacuum structure and the values of the various condensates
typically associated to the spontaneous breaking of some global symmetries.

Recently, a systematic approach has been devised \cite{dv,dglvz,cdsw}
to compute such low energy effective superpotentials as functions
of the glueball superfields, which are assumed to be the correct
low energy fields below the confining scale of the non-abelian gauge groups.
By virtue of the linearity principle and the holomorphy of such Wilsonian
effective superpotentials, the information obtained in this way is exactly
the same as the one obtained by integrating out the (generically massive)
glueball fields $S_i$. See \cite{review} for a review of this approach
and a list of references.

In this paper we will consider 
a supersymmetric ${\cal N}=1$ $U(N)$ gauge theory with
matter consisting only of a chiral supermultiplet $\chi^{ij}=\chi^{[ij]}$ 
in the antisymmetric
representation, and its charge conjugate $\tilde \chi_{ij}=\tilde \chi_{[ij]}$.
One can argue \cite{ags} that
in the large $N$ limit its effective dynamics should be the same as
the one of a gauge theory with the same $U(N)$ gauge group but with 
matter in the adjoint representation. We will indeed reproduce this
large $N$ behavior. Here we wish on the other hand to determine 
the effective superpotential at finite $N$, to see whether a strict
equivalence can be established, as for instance it was shown
in \cite{equiv} for the theory with fundamental matter.

We find that the theory with antisymmetric matter is actually more subtle,
and has subleading (in $N$) corrections which yield a superpotential
such that the $N$-dependence does {\em not} factorize. This behavior
is reminiscent of $SO/Sp$ theories with matter in the symmetric/antisymmetric
representation \cite{ks,ac,krs,cachazo,matone} 
(i.e. not in the adjoint). Actually we will see that not
only the $U(N)$ theory with antisymmetric matter can be solved in a 
very similar way, but there is actually a precise map between this theory
and an $Sp(\tilde N)$ theory with antisymmetric matter, with $\tilde N=
2N-2$ (we use the convention where $\tilde N$ is always even).

Note that in the theory we consider there is no field in the adjoint present. 
The theories
where matter both in the adjoint and in the (anti)symmetric representations
are present have been considered in \cite{kllr,nsw,klt}. We will discuss
the relation of those theories to the present one in the concluding 
discussion.

While preparing this paper for submission, we have received \cite{landsteiner}
where the same theory is considered. The approach is similar and the
conclusions are consistent, though the focus is slightly different. 

\section{Generalized Konishi anomaly}

We begin by reviewing the chiral ring of the $U(N)$ gauge theory with 
matter in the antisymmetric representation.
The group action on the present representation: 
\beqs
(\WW_\alpha^a T_a \chi)^{ij} & = & {{\WW_\alpha}^i}_k\chi^{kj} +
  {{\WW_\alpha}^j}_k\chi^{ik}, \nn \\
 & = & (\WW_\alpha \chi)^{ij} + (\chi \WW_\alpha^T),  
\eeqs
leads to the following relation in the chiral ring:
\beq
\WW_\alpha \chi = - \chi \WW_\alpha^T,
\eeq
and similarly for $\tilde \chi$, $\tilde \chi \WW_\alpha= - \WW_\alpha^T
\tilde\chi$, so that $\WW_\alpha$
commutes with the pair $\chi\tilde \chi$. Note that for future convenience
we write $\WW_\alpha$ in the fundamental representation, i.e. as $N\times N$
matrices.

As usual, non trivial operators in the chiral ring cannot have more
than two $\WW_\alpha$. On the other hand, arbitrary powers of  
$\chi$ and $\tilde \chi$ can be multiplied, provided that they are 
alternated, $\dots \chi^{ij}\tilde \chi_{jk}\chi^{kl}\tilde\chi_{lp}\dots$.
Hence the most general gauge-variant operators in the chiral ring are:
\beq
\WW^n (\chi \tilde \chi)^k, \quad \WW^n (\chi \tilde \chi)^k \chi,
\quad \tilde \chi \WW^n (\chi \tilde \chi)^k, \quad
\tilde \chi \WW^n (\chi \tilde \chi)^k \chi,
\eeq
with $n\leq 2$ and $k$ arbitrary.
It is much like the theory with the adjoint, in the sense that 
one can construct
independent polynomials in $\chi$ and $\tilde \chi$. Note that this is not
the case in the theory with the fundamental, where the meson
operator is the only independent composite operator in the chiral ring.

Gauge invariant operators in the chiral ring are just given by:
\beq
\tr \WW^n (\chi \tilde \chi)^k .
\eeq

We now take a generic tree level superpotential:
\beq
\wtree = \sum_{k=1}^{m} {1\over k} g_k \tr (\chi\tilde \chi)^k.
\eeq
Taking simply the equation for 
the anomalous $U(1)$ rotations of the superfield $\chi$,
we obtain:
\beq
\bar D^2(\chi^\dagger e^V \chi)= \sum_{k=1}^{m} g_k \tr (\chi\tilde \chi)^k
-(N-2) S, \label{simplekon}
\eeq
with as usual,
\beq
S=-{1\over 32 \pi^2} \tr \WW^\alpha \WW_\alpha = 
 {1\over 16  \pi^2} \tr \lambda^\alpha \lambda_\alpha + \dots.
\eeq
Since in eq.~\refe{simplekon}, as soon as $m\geq 2$, several independent 
chiral operators appear, it is not possible to solve for $\langle
\tr (\chi\tilde \chi)^k\rangle$ in
terms of $\langle S\rangle$ and $g_k$ using simply this equation
(as it was possible for the theory with the fundamental for instance).
We need to derive more relations, in the spirit of \cite{cdsw} for
the theory with the adjoint.

We will take more general variations, which read:
\beq
\delta \chi \equiv F  = \WW^n (\chi \tilde \chi)^p \chi .
\label{genvar}
\eeq
We will be only interested in the cases $n=0,2$, and will actually
consistently impose that the gauge invariant operators with one $\WW$
vanish. Note that for $n=0=p$ we recover the usual linear rotation.

For an arbitrary representation $\Phi^r$,
the general chiral ring relation following from the one-loop anomaly in
the variation $\delta \Phi^r=F^r(\WW,\Phi)$ reads:
\beq
{\partial \wtree(\Phi) \over \partial \Phi^r}F^r =  -{1\over 32 \pi^2}
{{\WW^\alpha}^r}_s{{\WW_\alpha}^s}_t{\partial F^t\over \partial \Phi^r}.
\eeq
Applying this general formula to the antisymmetric representation,
we obtain:
\beq
{\partial \wtree \over \partial \chi^{ij}}F^{ij}=
 -{1\over 32 \pi^2} 2 \left({{\WW^\alpha}^i}_j{{\WW_\alpha}^j}_k
{\partial F^{kl}\over \partial \chi^{il}}
+{{\WW^\alpha}^i}_j{{\WW_\alpha}^l}_k 
{\partial F^{jk}\over \partial \chi^{il}} \right). \label{genkon}
\eeq
Using \refe{genvar}, we obtain:
\beqs
{\partial F^{jk}\over \partial \chi^{il}}&=& {1\over 4} \sum_{r=0}^p 
\left\{{[\WW^n(\chi\tilde\chi)^r]^j}_i {[(\chi\tilde\chi)^{p-r}]^k}_l
-{[\WW^n(\chi\tilde\chi)^r]^k}_i {[(\chi\tilde\chi)^{p-r}]^j}_l\right. \nn \\
& & \left.\qquad
-{[\WW^n(\chi\tilde\chi)^r]^j}_l {[(\chi\tilde\chi)^{p-r}]^k}_i
+{[\WW^n(\chi\tilde\chi)^r]^k}_l {[(\chi\tilde\chi)^{p-r}]^j}_i\right\}.
\eeqs
For reference, we also write:
\beqs
{\partial F^{kl}\over \partial \chi^{il}}&=& {1\over 4} \sum_{r=0}^p 
\left\{\tr(\chi\tilde\chi)^{p-r} {[\WW^n(\chi\tilde\chi)^r]^k}_i
+\tr\WW^n(\chi\tilde\chi)^r {[(\chi\tilde\chi)^{p-r}]^k}_i\right\} \nn \\
& & \qquad \qquad - {1\over 2} (p+1) {[\WW^n(\chi\tilde\chi)^p]^k}_i.
\eeqs
As a consistency check, we note that taking $n=p=0$ we obtain
${\partial \chi^{jk}\over \partial \chi^{il}}={1\over 2}
(\delta^j_i \delta^k_l-\delta^k_i \delta^j_l)$ and
${\partial \chi^{kl}\over \partial \chi^{il}}={1\over 2}(N-1)\delta^k_i$,
and thus from \refe{genkon} we recover \refe{simplekon}.

With the general variation, we get:
$$ 
\sum_{k=1}^m g_k \tr\WW^n (\chi\tilde\chi)^{k+p} =  -{1\over 32 \pi^2}
\left[{1\over 2}\sum_{r=0}^p \left\{\tr(\chi\tilde\chi)^{p-r}
\tr\WW^{n+2}(\chi\tilde\chi)^r\right.\right. \hspace{2cm}$$ 
\beq\left.\left.\hspace{2cm} +
\tr\WW^n(\chi\tilde\chi)^r \tr\WW^2(\chi\tilde\chi)^{p-r}\right\}
-2(p+1)\tr\WW^{n+2}(\chi\tilde\chi)^p{\over}\right].
\eeq
We thus have two sets of equations, for $n=2$ and for $n=0$:
\beq
\sum_{k=1}^m g_k \tr\WW^2 (\chi\tilde\chi)^{k+p} = -{1\over 32 \pi^2}
{1\over 2} \sum_{r=0}^p\tr\WW^2(\chi\tilde\chi)^r 
\tr\WW^2(\chi\tilde\chi)^{p-r}, \label{rcomp}
\eeq
\beq
\sum_{k=1}^m g_k \tr (\chi\tilde\chi)^{k+p} = -{1\over 32 \pi^2}
\left[\sum_{r=0}^p\tr\WW^2(\chi\tilde\chi)^r \tr(\chi\tilde\chi)^{p-r}
-2(p+1)\tr\WW^2(\chi\tilde\chi)^p\right]. \label{tcomp}
\eeq
In order to obtain two closed equations for two generating
functions of gauge invariant operators (the resolvents), we multiply
both sides of the above equations by $z^{-2p-2}$ and sum over all $p\geq 0$.
We thus obtain from \refe{rcomp}:
\beq
\sum_{k=1}^m g_k \tr{\WW^2 (\chi\tilde\chi)^k\over z^2 -\chi\tilde\chi}
= -{1\over 32 \pi^2} {1\over 2} \left( z \tr{\WW^2\over z^2 -\chi\tilde\chi}
\right)^2.
\eeq
We thus define the resolvent:
\beq
R(z)= -{1\over 32 \pi^2}  z \tr{\WW^2\over z^2 -\chi\tilde\chi},
\eeq
which has the usual behavior $R\sim {S\over z}$ for large $z$.
Note a technical subtlety: in the adjoint case, we could write \refe{rcomp}
for $p=-1$ (or equivalently, for $\delta \Phi^i_j= \delta^i_j$), and that 
would correspond to a simple pole in $z$ on the left hand side. Here this
term is not present, and  this is why we have to be slightly more subtle
in the definition of $R(z)$.\footnote{In \cite{landsteiner}, this subtlety
is treated in a different, but consistent, way.} 

The equation for $R(z)$ reads:
\beqs
 {1\over 2}R(z)^2 & = &  -{1\over 32 \pi^2}
 \sum_{k=1}^m g_k \tr{\WW^2 (\chi\tilde\chi)^k\over z^2 -\chi\tilde\chi}\nn \\
&=& -{1\over 32 \pi^2}
\sum_{k=1}^m g_k \tr\WW^2 {(\chi\tilde\chi)^k-z^{2k} +z^{2k}
 \over z^2 -\chi\tilde\chi} \nn \\
&=& -{1\over 32 \pi^2} \sum_{k=1}^m g_k z^{2k-1} \, z \tr{\WW^2\over z^2 -
\chi\tilde\chi} -{1\over 32 \pi^2} 
\sum_{k=1}^m g_k \tr\WW^2 {(\chi\tilde\chi)^k-z^{2k}  
\over z^2 -\chi\tilde\chi} \nn \\
&=& R(z) W'(z) +  {1\over 2} f(z). \label{ranti}
\eeqs
We thus see that, since we are assigning a power of $z$ to every field, 
we need to define a superpotential function of degree $2m$, 
$W(z)=\sum_{k=1}^m {1\over 2k} g_k z^{2k}$. The polynomial $f(z)$ is 
in his turn of degree $2m-2$.

The solution to the above equation is:
\beq
R(z)= W'(z) - \sqrt{ W'(z)^2 + f(z)}\equiv W'(z) - y(z).
\eeq
Under the square root we have a polynomial of degree $4m-2$, so that
typically $y(z)$, and thus $R(z)$ will have $2m-1$ cuts on the complex
plane. Note that until now, the structure is very similar to the
$U(N)$ theory with adjoint matter and an even tree level superpotential.

We now sum the eqs.~\refe{tcomp}. We obtain:
\beq
\sum_{k=1}^m g_k \tr{ (\chi\tilde\chi)^k\over z^2 -\chi\tilde\chi} =
-{1\over 32 \pi^2} z \tr {1 \over z^2 -\chi\tilde\chi}
z  \tr{ \WW^2 \over z^2 -\chi\tilde\chi} 
-{1\over 32 \pi^2} z {d\over dz}  \tr{ \WW^2 \over z^2 -\chi\tilde\chi}.
\eeq
The second term on the r.h.s. derives from the single trace term in
the r.h.s of \refe{tcomp}.

Defining:
\beq
T(z) = z \tr {1 \over z^2 -\chi\tilde\chi},
\eeq
we obtain the following equation:
\beq
\sum_{k=1}^m g_k z^{2k-1} T(z) + \sum_{k=1}^m g_k \tr { (\chi\tilde\chi)^k
-z^{2k}\over z^2 -\chi\tilde\chi} = T(z) R(z) +  z {d\over dz} \left(
{1\over z}R(z) \right), 
\eeq
or:
\beq
W'(z) T(z) + c(z) = T(z) R(z) -{1\over z}R(z) + R'(z). \label{tanti}
\eeq
The leading order term of this equation, which is $1/z^2$,
reproduces the relation \refe{simplekon}. Note that unlike
in the adjoint case, we have no $1/z$ term, which in that case 
reproduced the (traced) classical equations of motion. Here 
simply the classical equations cannot be traced.

We now realize that the equation for $T(z)$ is rather different from 
the one for matter in the adjoint, because of the two additional
terms on the r.h.s. These two terms are reminiscent of those
that appear for $SO/Sp$ theories \cite{ac,krs} with matter in the adjoint
(the ${1\over z}R$ term) or in the symmetric/antisymmetric
(the $R'$ term). This analogy will be pushed further below.

For the time being, let us solve for $T(z)$, recalling that $y=W'-R$:
\beq
T(z)= {\tilde c(z) \over y(z)} + {1\over z} - {d\over dz}\log y(z),
\label{yanti}
\eeq
where we have redefined the polynomial of degree $2m-2$ to be
$\tilde c(z) = W''(z)-c(z)-{1\over z}W'(z)$.

Before analyzing the general case, let us pause for a moment and consider
the trivial case of $\wtree = m \tr\chi \tilde \chi$. Here the ordinary
Konishi anomaly is sufficient to solve for the effective superpotential, 
but let us solve for the resolvents, in order to obtain the VEVs of
all the gauge invariants in the SUSY vacuum.

Solving for $R(z)$, and fixing the constant $f$ by requiring that
for large $z$ we have $R\sim {S\over z}$, we obtain:
\beq
R(z) = m z -\sqrt{m^2 z^2 -2mS}.
\eeq
Also $T(z)$ is readily found, with $c$ determined by the large $z$ behavior
$T\sim {N\over z}$:
\beqs
T(z)&=&{N\over \sqrt{z^2-{2S\over m}}} - {z\over z^2-{2S\over m}}
+{1\over z} \\
&=& {N\over z}\left(1 + {S\over m z^2} +{3S^2\over2 m^2 z^4}+\dots\right)
+{1\over z}\left( - {2S\over m z^2} -{4S^2\over m^2 z^4}+\dots\right)
\nn
\eeqs
We thus find for the first two VEVs:
\beq
\tr\chi \tilde \chi = (N-2){S\over m}, \qquad 
\tr(\chi \tilde \chi)^2 = {3N-8 \over 2}{S^2\over m^2}.
\eeq
Contrary to what happens for the adjoint case, even in this simple
setting the VEVs fail to display a common $N$-dependence.

Let us clarify here a possible source of confusion. If one 
applies the above formulas to the limiting cases of $N=2$ or $N=3$, 
where the antisymmetric is, respectively, the singlet and the
(conjugate) fundamental, one quickly finds contradictions with the expected
results of vanishing condensates for $N=2$ and simple powers
of the meson condensate for $N=3$ (in the latter case the discrepancies arise 
at the next order). However, there is no contradiction, since
one is actually computing the VEVs of chiral operators which can classically
be expressed in terms of products of lower dimensional chiral operators.
Now, these chiral ring relations can, and do, get non-perturbative 
quantum corrections
(starting from the operator whose VEV is proportional to $S^N$,
i.e. to a one-instanton contribution). The analogous phenomenon
for the theory with adjoint matter is analyzed in \cite{cdsw}
(see also \cite{svrcek}). Note that as long as such higher order operators
do not appear in $\wtree$, $\weff$ as computed in the following
sections coincides strictly with the one computed, for instance 
for $N=3$, by replacing the antisymmetric fields with fundamental ones.

\section{Relation to an $SO$ theory with adjoint matter}

We are now going to see that, in order to correctly solve this theory,
we have to embed it into an $SO$ theory with adjoint matter, much
in the spirit of the treatment by \cite{cachazo} for the $Sp$
theory with antisymmetric matter.

Let us recall what the equations for the resolvent are in the
$SO$ theory with adjoint matter $\phi$ \cite{ac,krs}. 
We put a hat on $SO$ quantities.
Note that the tree level superpotential must be even because
of antisymmetry of $\phi$, $\wtree=\sum_1^m {1\over 2k}h_{2k} \tr \phi^{2k}$.
In terms of the resolvents:
\beqs
\hat R(z) &=&  - {1\over 32 \pi^2}z \tr {\WW^2 \over z^2 - \phi^2}, \\
\hat T(z) &=&  \tr {z \over  z^2 - \phi^2}, 
\eeqs
the two equations derived from the generalized Konishi anomalies read:
\beq
{1\over 2} \hat R(z)^2 = \hat R(z) \hat W'(z) + {1\over 2} \hat f(z),
\label{rso}
\eeq
\beq
\hat T(z) \hat R(z) - {2\over z}  \hat R(z)= 
\hat T(z) \hat W'(z) + \hat c(z), \label{tso}
\eeq
with $ \hat f(z)$ and $\hat c(z)$ two polynomials of degree $2m-2$ much
similar to $ f(z)$ and $ c(z)$.
We thus immediately see that \refe{rso} is the same as \refe{ranti},
while \refe{tso} features the $R/z$ term, present also in \refe{tanti}.

Note that since traces of odd powers of adjoint $SO$ matrices vanish, 
we could also write more familiarly:
\beq
\hat T(z) = \tr {z \over  z^2 - \phi^2} = {1\over 2}\left( 
\tr {1\over z-\phi} + \tr {1\over z+\phi}\right) = \tr {1\over z-\phi},
\eeq
by using $\phi^T=-\phi$, and similarly for $\hat R(z)$.
The $1/z^2$ terms of \refe{tso} lead to the ordinary Konishi anomaly
relation:
\beq
\sum_{k=1}^m h_{2k} \tr \phi^{2k} = (N-2) \hat S.
\eeq
Note that with this normalization the VY term will have a prefactor
of ${1\over 2}(N-2)$. For later convenience we prefer to keep this
normalization, although for purely $SO$ considerations it would be
preferable to rescale $\hat S$ to $2\hat S$.

Defining as before $\hat y=\hat W' - \hat R$, we can write the solution
for $\hat T$ as:
\beq
\hat T(z)= {\hat {\tilde c}(z)\over \hat y(z)} + {2\over z}, \label{yso}
\eeq
with $\hat {\tilde c}(z) = -\hat c(z) - {1\over z}\hat W'(z)$.
We thus observe that the structure of $\hat T(z)$ is the same
as the one for a $U(N)$ theory with adjoint matter, except
for an additional pole at the origin. Moreover, this additional
pole does not contribute at all to the expression for the VEVs
$\langle \tr \phi^{2k}\rangle$. Indeed, it is possible to show
that, for instance, in the one-cut case (classically unbroken $SO(N)$),
all the above VEVs are proportional to the respective VEVs in the
$U(N)$ theory, with $N-2$ factorizing in front of them instead of $N$.
The relation between $SO$ and $U$ theories with adjoint matter
has been discussed in \cite{achkr,jo,ikrsv}.

We are now going to embed the $U(N)$ theory with the antisymmetric
into an $SO(\hat N)$ theory with adjoint matter. 

First of all, comparing \refe{ranti} and \refe{rso}, we can just
equate $\hat R(z) = R(z)$, $\hat W'(z)= W'(z)$ and $\hat f(z) = f(z)$,
and thus $\hat y(z) = y(z)$. Note that this also implies $\hat S=S$.

On the other hand, comparing, for instance, \refe{yanti} and
\refe{yso}, we obtain a non-trivial relation:
\beq
\hat T(z) = 2 T(z) + {d\over dz} \log y(z)^2, \qquad \qquad \hat {\tilde c}(z)
=2 \tilde c(z).
\eeq
Remembering that $y^2(z)$ is a polynomial of degree $4m-2$, we can immediately
fix the relation between $\hat N$ and $N$:
\beq
\hat N = {1\over 2\pi i}\oint_{C_\infty} \hat T(z)dz={2\over 2\pi i}
\oint_{C_\infty}
  T(z)dz
+{1\over 2\pi i}\oint_{C_\infty} {d\over dz} \log y(z)^2 dz = 
2N + 4m -2,
\eeq
so that the $U(N)$ theory is mapped to a $SO(2N +4m-2)$ theory.

More importantly, we find that in the unbroken $U(N)$ vacuum, we
have:
\beq
\hat N_i=  {1\over 2\pi i}\oint_{C_i} \hat T(z)dz  =
{2\over 2\pi i}\oint_{C_i}  T(z)dz + {1\over 2\pi i} \oint_{C_i}
 {d\over dz} \log y(z)^2 dz = 2,
\eeq 
since ${d\over dz} \log y(z)^2$ has a pole at both edges of every cut
(or a pole of residue 2 if the cut factorizes into a simple zero), and
$C_i$ circles around the $i$th cut not containing the origin.

We thus conclude that in order to study the unbroken vacuum of the
$U(N)$ theory, we need to study the $SO$ theory with symmetry
breaking pattern:
\beq
SO(2N +4m-2) \rightarrow SO(2N+2)\times U(2)^{m-1}.
\eeq
More generally we expect a symmetry breaking pattern:
\beq
U(N)\rightarrow U(N_0)\times Sp(N_1) \times
\dots \times Sp(N_{m-1}),
\eeq
with $N=N_0+\sum_1^{m-1}N_i$, to map to:\footnote{One might be worried 
that the total number of vacua does not match. Indeed we have $2^m$ times
more vacua in the second theory. A similar controversy is discussed and 
solved in \cite{cachazo}. Here we effectively hide it with our normalization
of $\hat S$.}
\beq 
SO(2N +4m-2) \rightarrow SO(2N_0 +2) \times U(N_1+2) \times \dots \times
U(N_{m-1}+2).
\eeq

Much in the same way as in $Sp$ with antisymmetric matter \cite{cachazo}, 
even in the simple case of the classically unbroken gauge group, we need
to deal with a multicut solution. 
Note indeed that the unbroken gauge group scenario
can be thought of as actually displaying $Sp(0)$ factors, which have
been studied in \cite{ll,ikrsv}.
This is a first hint that the $N$-dependence
will not factorize in front of the effective superpotential, as indeed it
is the case in the above mentioned case \cite{ks,ac,krs}.

\section{Effective superpotential for a quartic interaction}

From now on, we will specialize to a theory with a quartic tree level 
superpotential:
\beq
W(z) = {1\over 2} m z^2 + {1\over 4} \lambda z^4.
\eeq
The mapping is thus to a $SO(2N+6)$ theory with symmetry breaking pattern
$SO(2N+6)  \rightarrow SO(2N+2)\times U(2)$.
The function $y(z)$ will thus have a cut around the origin and two
symmetric cuts around the classical extrema of the superpotential, 
$\pm\sqrt{-{m\over \lambda}}$.

We first want to compare the VEVs of the operators which appear 
in the tree level superpotential. They correspond to the $1/z^3$ and
$1/z^5$ terms in the expansions of $T(z)$ and $\hat T(z)$. What we
need to do is to find the corresponding terms in the expansion of
${d\over dz} \log y(z)^2$.
The leading behavior of $f$ is $f\sim -2\lambda S z^2$ (it is fixed imposing 
$R\sim S/z$ for large $z$), thus we find:
\beq
{d\over dz} \log y(z)^2 = {6\over z} - {4m \over \lambda z^3}
+ \left({4m^2 \over \lambda^2} + {8S\over \lambda}\right)  {1\over z^5}
+\dots,
\eeq
so that we find:
\beqs
\langle \tr \chi \tilde \chi\rangle &= & 
{1\over 2} \langle \tr \phi^2 \rangle +  {2m \over \lambda}, \\
\langle \tr( \chi \tilde \chi)^2 \rangle &= & 
{1\over 2} \langle \tr \phi^4 \rangle -  {2m^2 \over \lambda^2}
- {4S\over \lambda}. \label{quarticrel}
\eeqs
The corrections independent of $S$ take care of the classical
part of the VEVs (which is present in the $SO$ theory but
absent in the $U(N)$ theory), while the correction linear
in $S$ is actually related to the matching of the scales of the two
theories, as we discuss below.

What we are left to do is a quite laborious procedure. We should
solve the $SO(2N+6)$ theory in terms of the two glueball superfields
corresponding to the two low-energy gauge groups, extremize its
effective superpotential, relate the latter to the $U(N)$ 
effective superpotential
through the relations above (and the relation between the holomorphic scales),
and eventually integrate in the glueball superfield of the $U(N)$ theory.

Instead, we will begin with a discussion on the scales of the theories
involved, which clarifies when the subtleties related to the effective
multi-cut solution set in (see \cite{ikrsv} for a similar discussion
in the $Sp$ context).

First of all, let us see how \refe{quarticrel} determines the matching
of the scale $\Lambda_h$ of the $U(N)$ theory with the scale $\hat \Lambda_h$
of the $SO(2N+6)$. These are both high energy scales, related
to the beta function of the theories with matter present.
From \refe{quarticrel} we can determine:
\beq
\weff^U = \weff^{SO} -2S \log \lambda \mu + {m^2 \over \lambda}, 
\label{usorel}
\eeq
where $\mu$ is, say, the renormalization scale.
Let us now consider that the scale dependent piece of $\weff$
reads, in general:
\beq
\weff = - \beta_0 S \log {\Lambda_h \over \mu}.
\eeq
This is basically the tree-level plus one-loop contribution to the
superpotential, subtracted in order to write the effective superpotential
in terms of $S$ (i.e. after integrating it in).
Thus the relation between the scales becomes:
\beq
-(2N+2)S \log {\Lambda_h \over \mu} = 
- (2N+4) S \log {\hat \Lambda_h \over \mu}  -2S \log \lambda \mu ,
\eeq
where we have used that $\beta_0^U= 3N- (N-2)$ and $\beta_0^{SO}=
2 (2N+4)$, paying attention to the normalization we use on the $SO$
side. We find the relation:
\beq
\Lambda_h^{2N+2}=\hat \Lambda_h^{2N+4}\lambda^2.
\eeq
We now want to derive another relation, 
expressing the two low-energy scales $\hat \Lambda_0$ and $\hat 
\Lambda_1$ 
of the $SO$ theory in terms of the unique low-energy scale $\Lambda$
of the $U(N)$ theory.
In order to do this, we have to match the scales using the mass $m$ of
the matter field and, on the $SO$ side, the VEV $|\phi|^2={m\over \lambda}$
inducing gauge symmetry
breaking (we will neglect all numerical factors in the following).

In the $U(N)$ theory, the matching is straightforward:
\beq
\Lambda^{3N} = \Lambda_h^{2N+2} m^{N-2}.
\eeq
In the $SO(2N+6)$ theory, we have to introduce the intermediate scales
$\hat \Lambda_{0,int}$ and $\hat \Lambda_{1,int}$ of the $SO(2N+2)$
and $U(2)$ theories with the adjoint field, respectively:
\beq
\hat \Lambda_{0,int}^{4N}=\hat \Lambda_h^{4N+8} {\lambda^4 \over m^4},
\qquad 
\hat \Lambda_{1,int}^4= \hat \Lambda_h^{4N+8}  {\lambda^{2N+2} 
\over m^{2N+2}}.
\eeq
From those, we can get the relation to the scales at low energies:
\beq
\hat \Lambda_0^{6N}=\hat \Lambda_{0,int}^{4N} m^{2N}=
\hat \Lambda_h^{4N+8} \lambda^4 m^{2N-4} = 
\Lambda_h^{4N+4} m^{2N-4} = \Lambda^{6N},
\eeq
\beq
\hat \Lambda_1^{6}=\hat \Lambda_{1,int}^4 m^2 =
 \hat \Lambda_h^{4N+8} {\lambda^{2N+2}  \over m^{2N}}
=\Lambda_h^{4N+4}  {\lambda^{2N-2}  \over m^{2N}}
=\Lambda^{6N} {\lambda^{2N-2}  \over m^{4N-4}}.
\eeq
We thus find that $\hat \Lambda_0 = \Lambda$, while $\hat \Lambda_1^{3}
\propto \Lambda^{3N}$, i.e. in the effective superpotential the first
contribution originally coming from the effective gauge dynamics in the $U(2)$
factor looks like a one-instanton contribution in the $U(N)$ effective
theory. In other words, the effects related to the fact that $y(z)$
does not really factorize and 
has three cuts instead of one will appear in $\weff(S)$ only
at order $S^N$ and beyond. 
This is much like in the $Sp$ with antisymmetric case, where the 
``discrepancies'' between the field theory expectations and the naive
matrix model computation set in at order $S^h$ \cite{ks}.

At this point, we could proceed to use the above information
to get, for instance, the effective superpotential $\weff(\Lambda,
m , \lambda)$ of the $U(N)$ theory in the following way:
first of all write the function $y$ in terms of 3 parameters;
relate those parameters to ${m\over \lambda}$, $S_0$ and $S_1$
by using for the latter two their definitions in terms of contour
integrals; write the expression for $\hat T(z)$, and determine
the coefficients of $\hat{\tilde c}(z)$ by imposing the residues
around the cuts corresponding to the size of the classical gauge
groups; expand $\hat T(z)$ to find the expressions for $\langle
\tr \phi^2\rangle$ and $\langle \tr \phi^4\rangle$; integrate
to find an expression\footnote{
Alternatively, we could obtain $\weff^{SO}$ by going through
the free energy of the associated matrix model, obtained by expanding
$R(z)$ rather than $\hat T(z)$.} 
for $\weff^{SO}$, to which one adds the
relevant VY pieces; extremize this expression
with respect to $S_0$ and $S_1$; substitute the expressions
for $\hat \Lambda_0$ and $\hat \Lambda_1$ in terms of $\Lambda$, and
eventually obtain $\weff^{U}$ using \refe{usorel}.

Another option, if we are satisfied in dealing case by case, 
is to analyze the factorization of the Seiberg-Witten curve,
which yields the result for, say,  $\langle \tr \phi^2\rangle$
directly in terms of $\hat \Lambda_h$. This is the route effectively
employed in \cite{cachazo}. At this effect, one could use the
results of \cite{ao} (see also \cite{factorization}).\footnote{
Note that a direct computation of the effective superpotential for $U(N)$
with antisymmetric matter
for low enough $N$ could also be done along the lines of \cite{ck,ks}, 
using the results on s-confining $SU(N)$ theories with antisymmetric
tensors and fundamentals found in \cite{css}.}

Here our main intention is to comment on the $N$-behavior of
the effective superpotential. We can thus content ourselves in
deriving the $N$-dependence for the first few terms in the $S$ 
expansion, and consider that $N$ is sufficiently large so that
the subtleties associated to the non-factorization of $y$
have no influence. In other words, as far as we are concerned
we can just blindly consider the $U(N)$ theory and suppose
that $y$ factorizes to a one-cut solution.\footnote{From the $SO$ 
perspective, this amounts to freezing the dynamics of the $U(2)$ 
factor by taking $S_1=0$, which implies $\hat \Lambda_1=0$.
}
We are actually going to argue that the VEVs have the same
expression as the VEVs of the $Sp$ theory with antisymmetric matter
and an even tree level superpotential (in the unbroken phase).
Since it is known explicitly that in the latter case the $N$-dependence
does not factorize, we thus conclude that also in the $U(N)$ theory
with antisymmetric matter the $N$-dependence does not factorize.
Of course, in the large $N$-limit, the leading $N$ behavior will
reproduce the effective superpotential for the adjoint matter.

Let us thus compute $\weff(S)$ at orders lower than $S^N$ by supposing
that the function $y(z)$ factorizes in the following way:
\beq
y(z)= \lambda (z^2-a^2)\sqrt{z^2-b^2}.
\eeq
The relation between the constants $a$ and $b$ to the other data of
the problem are obtained by remembering that $y^2=W'^2+f$, with
$f=-2\lambda S z^2 +f_0$. We obtain:
\beq
{m\over \lambda}= -a^2 -{1\over 2}b^2, \qquad \qquad
2{S\over \lambda} = {1\over 4} b^2 -a^2 b^2.
\eeq
Note that the second expression can alternatively be obtained
by performing a contour integral of $R(z)$ around the cut.
These relations can be inverted to:
\beqs
a^2 & = & -{m\over 3\lambda}\left( 2 + \sqrt{1+6{\lambda S\over
      m^2}}\right),\label{aexpr}\\
b^2 & = & {2m\over 3\lambda}\left( -1 + \sqrt{1+6{\lambda S\over
      m^2}}\right).\label{bexpr}
\eeqs
Correctly, for small ${\lambda S\over  m^2}$, we recover 
$a^2\sim -{m\over \lambda}$ and $b^2\sim 2 {S\over m}$.

We can now write:
\beq
T(z)= {c_2 z^2 + c_0 \over  (z^2-a^2)\sqrt{z^2-b^2}}
+{1\over z} - {1\over z-a }- {1\over z+ a }
-{1\over 2} {1\over z-b }-{1\over 2} {1\over z+b }.
\eeq
The first coefficient $c_2$ is easily fixed by asking that:
\beq
{1\over 2\pi i} \oint_{C_\infty}T(z)dz = c_2 -2 = N.
\eeq 
The second coefficient $c_0$ is then fixed requiring that 
the contour integral around the cut also gives $N$ or, alternatively,
that the contour integrals around the poles yield zero. This second
route is the most straightforward:
\beq
{1\over 2\pi i} \oint_{C_a}T(z)dz = {(N+2)a^2 +c_0 \over
2a \sqrt{a^2-b^2}} -1 = 0.
\eeq
We eventually end up with the following generating function:
\beq
T(z)= {N+2 \over \sqrt{z^2-b^2}} - {2\over z} -{b^2 \over z(z^2-b^2)}
+{2a \over z^2-a^2}\left({\sqrt{a^2-b^2} \over \sqrt{z^2-b^2}}
-{a\over z}\right). \label{tsol}
\eeq
We immediately see that the first term has the same structure
of the generating function for VEVs in the $U(N)$ with adjoint case, 
the second term only corrects the leading term giving the trace of
the identity, and most importantly the remaining terms will give
corrections which are $N$-independent and will be different at every
order in $1/z$.
Note that in the large $N$ limit, at leading order, we recover
\beq
T(z)\sim {N\over \sqrt{z^2-b^2}},
\eeq
which is exactly the generating function for $U(N)$ with adjoint matter.
In this sense for large $N$ and in the planar limit the effective
superpotentials for the two theories will coincide\footnote{
In the large $N$ limit, we can consistently work in the present
one-cut approximation, since the corrections at order $S^N$ are pushed
to infinity.}, as argued in \cite{ags}.
However we will see that the subleading corrections are more subtle
than argued there, since no $N$-dependence can be factorized at finite $N$.

By expanding the expression \refe{tsol}, we recover the VEVs:
\beqs
\langle \tr (\chi \tilde \chi)\rangle & = & 
{N\over 2} b^2 + 2a \sqrt{a^2-b^2} - 2a^2, \label{quadexpr}\\
\langle \tr (\chi \tilde \chi)^2 \rangle & = &
{3N-2\over 8} b^4 + 2a^3 \sqrt{a^2-b^2} - 2a^4 +ab^2  \sqrt{a^2-b^2}.
\eeqs
Using the expressions \refe{aexpr} and \refe{bexpr}, one can
rewrite the above VEVs exactly in terms of $S$, $m$ and $\lambda$,
and in principle it is possible to obtain by integration an exact
expression for $\weff(S, m, \lambda)$. We will refrain from doing so
here, also because the exact expression obtained in this way should
not be trusted to all orders but only up to $S^{N-1}$. Here we will
only compute the first few terms in the expansion in $S$, in order
to display their $N$-dependence.

For instance, since $b^2\propto S$, we can expand \refe{quadexpr}
in $b^2$:
\beq
\langle \tr (\chi \tilde \chi)\rangle = {N-2 \over 2} b^2 
-{1\over 4} {b^4 \over a^2} - {1\over 8} {b^6 \over a^4}+\dots.
\eeq
Using then the expansions:
\beq
b^2 = {2S\over m}\left[ 1 - {3\over 2} {\lambda S\over m^2} 
+ {9\over 2} \left({\lambda S\over m^2} \right)^2 +\dots \right], \qquad
a^2 = -{m\over \lambda } \left[ 1 +  {\lambda S\over m^2} +\dots \right],
\eeq
we find:
\beq
\langle \tr (\chi \tilde \chi)\rangle = (N-2) {S\over m}
-{1\over 2}(3N-8) {\lambda S^2\over m^3} + {1\over 2}(9N-28)
{\lambda^2 S^3\over m^5}+\dots . \label{weffexp}
\eeq
We thus obtain for the effective superpotential:
\beq
\weff = (N-2) S \log{m\over \mu} + {1\over 4}(3N-8) {\lambda S^2\over m^2}
-{1\over 8}(9N-28) {\lambda^2 S^3\over m^4} +\dots . \label{weffanti}
\eeq
The same expression, except the first term, can be obtained starting
from $\langle \tr (\chi \tilde \chi)^2\rangle$.

A quick look at \refe{weffanti} shows that the $N$-dependence does not
factorize. The $(N-2)$ behavior is restricted to the term responsible
for the threshold matching. 

Let us see what happens in two different limiting cases. 
First, consider the large $N$ limit. We can define the VEVs:
\beq
v_2 = {1\over N} \langle \tr (\chi \tilde \chi)\rangle
\simeq {1\over 2} b^2,
\eeq
\beq
v_4 = {1\over N} \langle \tr (\chi \tilde \chi)^2\rangle
\simeq {3\over 8} b^4 = {3\over 2} v_2^2.
\eeq
As in \cite{equiv}, we can insert these two expressions in the 
relation following from the ordinary Konishi anomaly:
\beq
m v_2 + {3\over 2}\lambda v_2^2 = S, \qquad \qquad \mbox{(large $N$)}
\eeq
so that, referring to the equivalent VEVs in the $U(N)$ theory with
adjoint matter, we have $v_2= u_2({S\over 2},m, \lambda)$, and
thus:
\beq
\weff^\mathrm{anti}(S,m,\lambda) = 2 \weff^\mathrm{adj} 
({S\over 2},m, \lambda).
\eeq
Hence, the effective superpotential for the $U(N)$ theory with the
antisymmetric is indeed essentially equivalent, in the leading large $N$
limit, to the one for the theory with the adjoint and
an even tree level superpotential, as argued in \cite{ags}.

The other limiting case we can check is when $N=3$. In this case, 
we know that for $U(3)$ the antisymmetric is actually the conjugate
fundamental, so that the theory boils down to a familiar case, $U(3)$
with a single flavor.
However, in this case we can trust our expansion \refe{weffexp}
only to order $S^2$, and we expect discrepancies at order $S^3$
and higher.

For $N=3$, \refe{weffexp} reads:
\beq
\weff^{N=3}=  S \log{m\over \mu} + {1\over 4}{\lambda S^2\over m^2}
+{1\over 8} {\lambda^2 S^3\over m^4} +\dots \label{weffthree}
\eeq
In order to compare with the known, exact, effective superpotential
for the theory with flavors, we first need to compare the VEVs.
It turns out that if we define $\chi^{ij}={1\over \sqrt{2}} \epsilon^{ijk}
\tilde Q_k$ and $\tilde \chi_{ij}=-{1\over \sqrt{2}} \epsilon_{ijk}  Q^k$,
we have that $\tr\chi \tilde \chi = Q \tilde Q \equiv X$ and
 $\tr(\chi \tilde \chi)^2={1\over 2}X^2$. Thus the relation following
from the Konishi anomaly reads:
\beq
mX+{\lambda \over 2} X^2 =S.
\eeq
Expanding now the expression found for instance in \cite{acfh,binor}
and replacing ${\lambda \over 2}$ for the quartic coupling, we get:
\beq
\weff^\mathrm{fund} = S\log{m\over \mu} + {1\over 4}{\lambda S^2\over m^2}
-{1\over 8} {\lambda^2 S^3\over m^4} +\dots.
\eeq
We thus see that the $S^2$ term is indeed reproduced by \refe{weffthree},
but, as expected, at order $S^3$ \refe{weffthree} starts to disagree with
the exact expression above.

\section{Relation to $Sp(\tilde N)$ with antisymmetric matter}

Let us comment on the relation between effective superpotentials
of different theories. The approach used here was reminiscent of
the one used for the $Sp(N)$ gauge theory with matter in the (traceful)
antisymmetric representation.
We now point out that the two effective superpotentials actually coincide,
upon replacing $N$ here with ${N\over 2}+1$ on the $Sp(N)$ side.
Note that on the $Sp(N)$ side an even classical superpotential
is not the most general one. Here for simplicity we only consider
a quartic $\wtree$.

First of all, let us note that we embed our $U(N)$ theory in a 
$SO(2N+6)$ theory with adjoint matter, while the $Sp(\tilde N)$ theory is
embedded in a $U(\tilde N+6)$ theory. The superpotential in a generic phase
of the $SO(2N+6)$ theory can be directly extracted from the superpotential
of the $U(\tilde N+6)$ theory \cite{ikrsv}, 
by taking in the latter the two cuts not
containing the origin to be symmetric, and identifying the two glueball
fields associated to them. After that, the expressions for the VEVs are
the same (and thus the effective superpotential)  since the additional
term in $T(z)$, see eq.~\refe{yso}, only affects the leading term
in the expansion. The only redefinition in $\weff$ is to map
the two dual Coxeter numbers, implying $2N_0-2=\tilde N_0$ for the
first factors ($SO(2N_0)$ and $U(\tilde N_0)$ respectively)
and $N_1=\tilde N_1=\tilde N_2$ for the other factors (which are all unitary).
In particular, when we have the breaking pattern $SO(2N+6)\rightarrow
SO(2N+2)\times U(2)$, on the other side we have $U(\tilde N +6)
\rightarrow U(\tilde N+2)\times U(2)\times U(2)$, with the glueballs
of the last two factors identified. The relation is thus $N={\tilde N\over 2}
+1$, that is the dual Coxeter number of $U(N)$ mapped to the one
of $Sp(\tilde N)$.

The logarithmic correction to the generating functions $T(z)$ is also the same
in the two cases under considerations. We thus conclude that the effective
superpotential is the same, up to the identification between the
dual Coxeter numbers. 

Note the following subtlety. It is quite straightforward to understand
that all terms up to $S^{h-1}$ share the same numerical factors, based
on the naive procedure outlined in the previous section. 
It is on the other hand less obvious that the ${\cal O}(S^h)$
corrections give also the same contributions. Indeed, 
on the $Sp(\tilde N)$ side we have contributions from two additional
$U(2)$ gauge groups, while on the $U(N)$ side the dynamics of only one $U(2)$
contributes, basically the diagonal one. 
However we are confident that upon extremization, 
the final expressions in terms
of the scale of the original theory with classically unbroken gauge group 
indeed coincide. For instance, 
a direct comparison using the $U(3)$ and $Sp(4)$
theories is possible. Quite straightforwardly one can reproduce the
exact superpotential for $Sp(4)$ with a quartic interaction\footnote{
The $Sp(4)$ theory
with antisymmetric matter has been solved in \cite{ks,cachazo}, though
only in the (actually more complicated) case where the classical 
superpotential is cubic.}
along the lines of \cite{ck,ks}, and it coincides with the one
for $U(3)$ with one flavor, even before integrating out the meson
superfield.

\section{Discussion}
We have shown in this paper that the effective superpotential
for a $U(N)$ theory with antisymmetric matter has an $N$-dependence
that does not factorize. Thus it does not share the same 
``universal'' functional form as the effective superpotential 
for $U(N)$ with adjoint matter, a functional form that was also 
shown to be reproduced by a $U(N)$ theory with fundamental matter
\cite{equiv}. On the other hand, we have shown that, for every $N$,
the superpotential is the same as the one arising in a theory
with $Sp(2N-2)$ gauge group and antisymmetric matter. Since the $N$-dependence
does not factorize, these superpotentials are different functions
of the glueball and couplings for every $N$.

The above results concern the vacuum with classically unbroken gauge group.
It is amusing to note the following thing. In vacua where on the
other hand the gauge group is classically maximally broken, 
$U(N)\rightarrow U(N_0)\times Sp(N_1) \times\dots \times Sp(N_{m-1})$, the 
structure is expected to be much more regular. In particular, the structure
of the effective superpotential (in terms of the glueballs $S_i$)
will be a sum of terms, each of which displaying a factorized
dependence on $N_i$.\footnote{More precisely, $N_0$ factorizes for
the $U(N_0)$ factors and ${N_i\over 2} +1$ factorizes for the $Sp$ factors.} 
This is because in this case the (in)famous
$Sp(0)$ factors are absent. Of course, all factorized dependence on $N_i$
is lost when the glueballs are integrated out. See \cite{ikrsv}
for the related $Sp$ case.

We cannot refrain from speculating that the equivalences among
effective superpotentials of theories differing by the gauge group
and/or the matter content might be due to underlying duality relations.
Also, it is intriguing that in all the cases considered, it is possible
by a chain of mappings to relate the theory under consideration to a
$U(N)$ theory with adjoint matter (with some restrictions on the
classical superpotential and symmetry breaking pattern). It is possible
that the generic solution to this latter theory contains all the information
to solve for all other theories, for classical groups and up to two
index representations at least. 

We have mentioned in the introduction that $U(N)$ theories with
both adjoint and antisymmetric matter have been analyzed \cite{kllr,nsw,klt}.
These theories have the special feature that they are solved in terms
of a cubic curve $y^3(z)=\dots$. 
While in our case the solution had a more traditional
quadratic curve associated to it, it should in principle be possible
to relate at least some of the results in both theories. Indeed, consider
the theory with the adjoint and a simple tree level superpotential like:
\beq
\wtree= {1\over 2} M \tr \Phi^2 + m \tr \chi \tilde \chi
+ g \tr \Phi \chi \tilde \chi.
\eeq
This is the simplest case of \cite{kllr,nsw,klt}, 
where a generic superpotential
for $\Phi$ is considered. However it should be exactly equivalent
to our case with a quartic superpotential, by classically integrating
out $\Phi$ (by holomorphy, we can always take $M$ to be very large in
this step) and identifying $\lambda\equiv -{g^2\over M}$. It would be
interesting to check that the superpotentials indeed coincide, as it
is the case for the theories with fundamental matter \cite{acfh}.

Finally, let us comment on a seemingly straightforward generalization,
that is to $U(N)$ with symmetric matter. Here we expect the theory
to be mapped to a $SO$ theory with symmetric matter. However, this
theory in the unbroken phase cannot be solved in exactly the same
way as the companion $Sp$ theory. Indeed, the trivial factors which
arise are $SO(0)$ factors, and those cannot be dealt with in the same
way as the $Sp(0)$ ones. See \cite{ikrsv} for a discussion on how
to deal with this, based on geometric transitions. It would be nice
to have an understanding of this case in pure gauge theoretic terms.

\subsection*{Acknowledgments}
I would like to thank A.~Armoni and G.~Ferretti for interesting discussions
and correspondence.
This work is supported in part by the ``Actions de Recherche
Concert{\'e}es" of the ``Direction de la Recherche Scientifique -
Communaut{\'e} Fran{\c c}aise de Belgique", by a ``P\^ole
d'Attraction Interuniversitaire" (Belgium), by IISN-Belgium
(convention 4.4505.86)  and by the European Commission RTN programme
HPRN-CT-00131. The author is a Postdoctoral Researcher of
the Fonds National de la Recherche Scientifique (Belgium).

\end{document}